\documentclass[a4paper]{article}
\usepackage{graphics}
\begin{document}

\parindent0pt

\centerline{\huge Quantum Phase Transition of $S=1/2$ Trimerized}
\bigskip
\centerline{\huge $XXZ$ Spin Chain in Magnetic Field}

\bigskip
\bigskip

\centerline{\Large ${}^{ \rm a}$Kiyomi Okamoto and ${}^{\rm b}$Atsuhiro Kitazawa}

\bigskip

\large

\centerline{\large ${}^{\rm a}$Department of Physics, Tokyo Institute of Technology,}

\centerline{\large Oh-okayama, Meguro-ku, Tokyo 152-8551, Japan}

\centerline{and}

\centerline{\large ${}^{\rm b}$Department of Physics, Kyushu University,}

\centerline{\large Hakozaki, Higashi-ku, Fukuoka 812-8581, Japan}

\normalsize

\bigskip
\bigskip

We study the magnetization plateau at a third of the saturation magnetization
of the $S=1/2$ trimerized $XXZ$ spin chain at $T=0$. 
The appearance of the plateau depends on the values of the $XXZ$ anisotropy
and the magnitude of the trimerization.
This plateauful-plateauless transition is a quantum phase transition of the
Berezinskii-Kosterlitz-Thouless type, which is difficult to precisely detect
from the numerical data. 
To determine the phase boundary line of this transition precisely,
we use the level crossing of low-lying excitations obtained from the numerical 
diagonalization. 
We also discuss the $S=1/2$ ferromagnetic-ferromagnetic-antiferromagnetic chain.

\medskip


Email address: kokamoto@stat.phys.titech.ac.jp

\newcommand{\vS}{{\mathbf S}}

\newpage

We study the magnetization plateau at $M=M_{\rm s}/3$
($M_{\rm s}$ is the saturation magnetization) of the
$S=1/2$ trimerized $XXZ$ spin chain described by [1]
\begin{eqnarray}
  H &=& \sum_{j=1}^{L}\left\{
  J^{'}\left[ h_{3j-2,3j-1}(\Delta)+h_{3j-1,3j}(\Delta) \right] \right.\nonumber \\
  & &~~~~~~\left. +J h_{3j,3j+1}(\Delta)\right\}
\end{eqnarray}
where
\begin{equation}
     h_{lm}(\Delta) = h^\perp_{lm} + \Delta h^z_{lm}
\end{equation}
\begin{equation}     
     h^\perp_{lm} = S^{x}_{l}S^{x}_{m}+S^{y}_{l}S^{y}_{m},~~~
     h^z_{lm} = S^{z}_{l}S^{z}_{m}.
\end{equation}
It is convenient to parametrize the Hamiltonian (1) as
\begin{equation}
    J=J_0(1+2t),~~~~~J'=J_0(1-t)
\end{equation}
where $-1/2 \le t \le 1$ is the trimerization parameter.
The bosonized expression of the Hamiltonian (4) 
has the sine-Gordon form
\begin{eqnarray}
  H
  &=& \frac{1}{2\pi} \int dx\left[ v_{\rm s}K(\pi\Pi)^{2}
    + \frac{v_{\rm s}}{K}
   \left(\frac{\partial \phi}{\partial x}\right)^{2}\right] \nonumber \\
  & &~~~~~~ + \frac{y_{\phi}v_{\rm s}}{2\pi}\int d x\cos\sqrt{2}\phi
\end{eqnarray}
where $v_{\rm s}$ is the spin wave velocity of the system,
$\Pi$ is the momentum density conjugate to $\phi$,
$[\phi(x),\Pi(x')] = i\delta(x-x')$, and
the coefficients $v_{\rm s}$, $K$, and $y_{\phi}$ are smooth functions of
$J_{0}$, $t$ and $\Delta$
The field $\phi$ is related to the fast varying (in space) part
of the spin density $S^z(x)$ in the continuum picture as
\begin{equation}
    S^z_{\rm fast}(x)
    = {1 \over3}
      \left\{ \cos \left(2k_{\rm F}x - {\pi \over 3}+ \sqrt{2}\phi \right)
              + {1 \over 2} \right\}
\end{equation}
which makes it clear the physical meaning of $\phi$.

~~~As is well known, the excitation spectrum of the sine-Gordon model is
either massive (plateauful)  or massless (plateauless)
depending on the values of parameters.
These two states are distinguished by the renormalized value of $K$:
the former is realized when $K<4$ and the latter $K\ge 4$.
The transition between these two states is of the Brerezinskii-Kosterlitz-Thouless
type, which was first pointed out by one of the present authors (K.O) [2].
To determine the $K=4$ BKT transition point from the numerical diagonalization data
for finite systems, we use the method
developed by Nomura and Kitazawa [3].
They discussed the $K=4$ BKT transition at $M=0$ to find that
the crossing between the $M=0$ excitation with the twisted boundary condition (TBC)
and the $M=2$ excitation with the periodic boundary condition (PBC). 
Because we discuss the finite magnetization case now, 
it is necessary to use the Legendre transformation $E \to E - HM$.
Thus we can conclude that the crossing between 
\begin{equation}
     \Delta E^{\rm TBC}
     = E_0^{\rm TBC} \left(\frac{M_{\rm s}}{3}\right)
      -E_0^{\rm PBC} \left(\frac{M_{\rm s}}{3}\right)
\end{equation}
and 
\begin{equation}
    \Delta E^{\rm PBC}
    =  \tilde E_0^{\rm PBC} 
             -E_0^{\rm PBC} \left( {M_{\rm s} \over 3} \right)\,.
\end{equation}
where $\tilde E_0^{\rm PBC} = (1/2)\{ E_0^{\rm PBC}(M_{\rm s}/3+2) 
+ E_0^{\rm PBC}(M_{\rm s}/3-2) \}$.
Figure 1 shows the crossing between these two excitations, from which we see
$\Delta_{\rm c}(L=18)=-0.8389$.
By sweeping parameters, 
we obtain the phase diagram on the $\Delta - t$ plane (Fig.1).

\begin{figure}[h]
   \hskip0.05\linewidth
   \begin{minipage}{.4\linewidth}
      \begin{center}
         \scalebox{0.3}[0.3]{\includegraphics{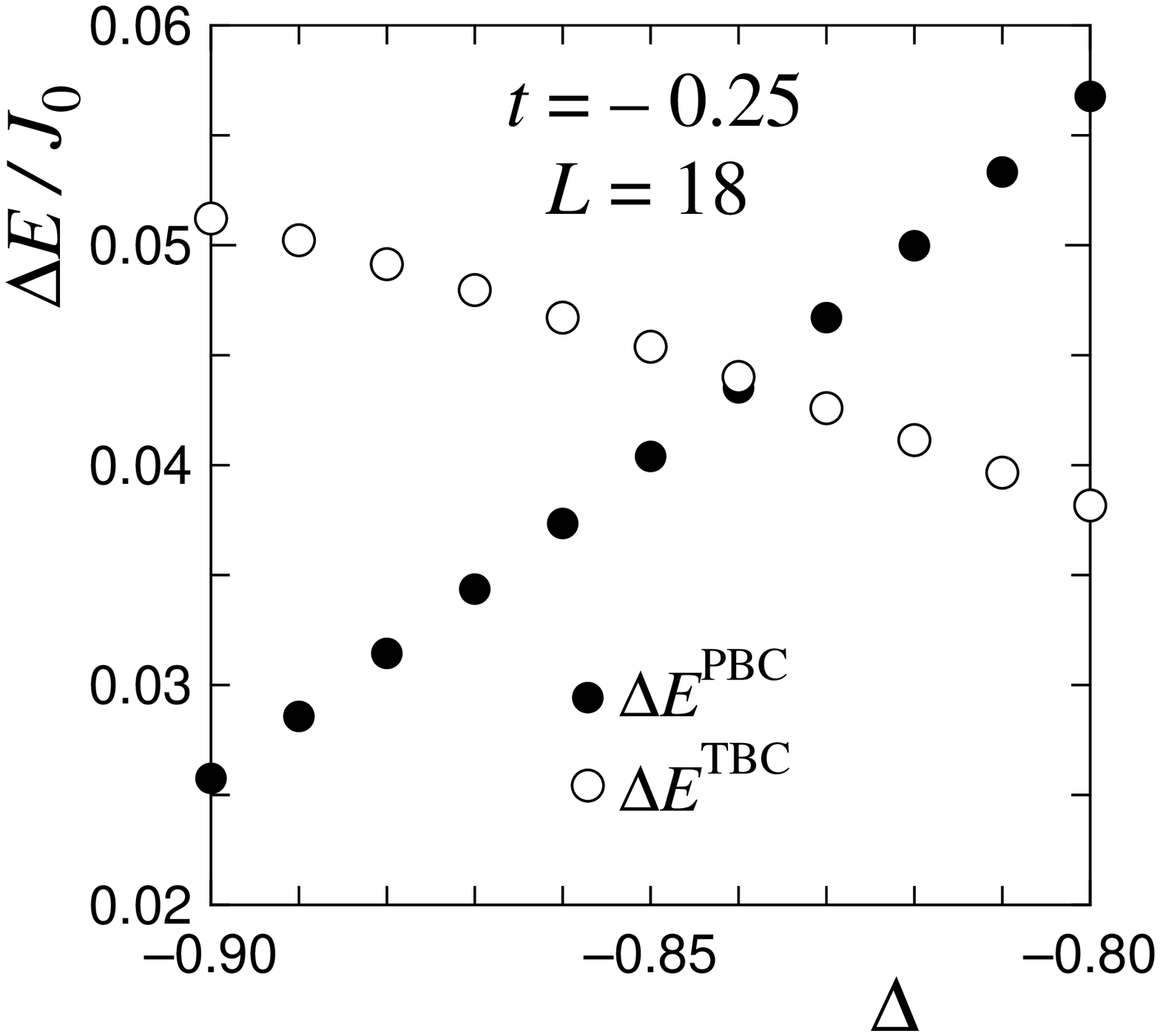}}
      \end{center}
      \caption{$\Delta E^{\rm TBC}$ (opne circles) and $\Delta E^{\rm PBC}$
         (closed circles) for $L=18$ spins
         as functions of anisotropy parameter $\Delta$ when $t=-0.25$.
         From the crossing point we obtain $\Delta_{\rm c}(L=18)=-0.8389$.}

      ~
      
      ~
      
      \label{fig:level-cross}
   \end{minipage}
   \hskip0.1\linewidth
   \begin{minipage}{.4\linewidth}
      \begin{center}
         \scalebox{0.3}[0.3]{\includegraphics{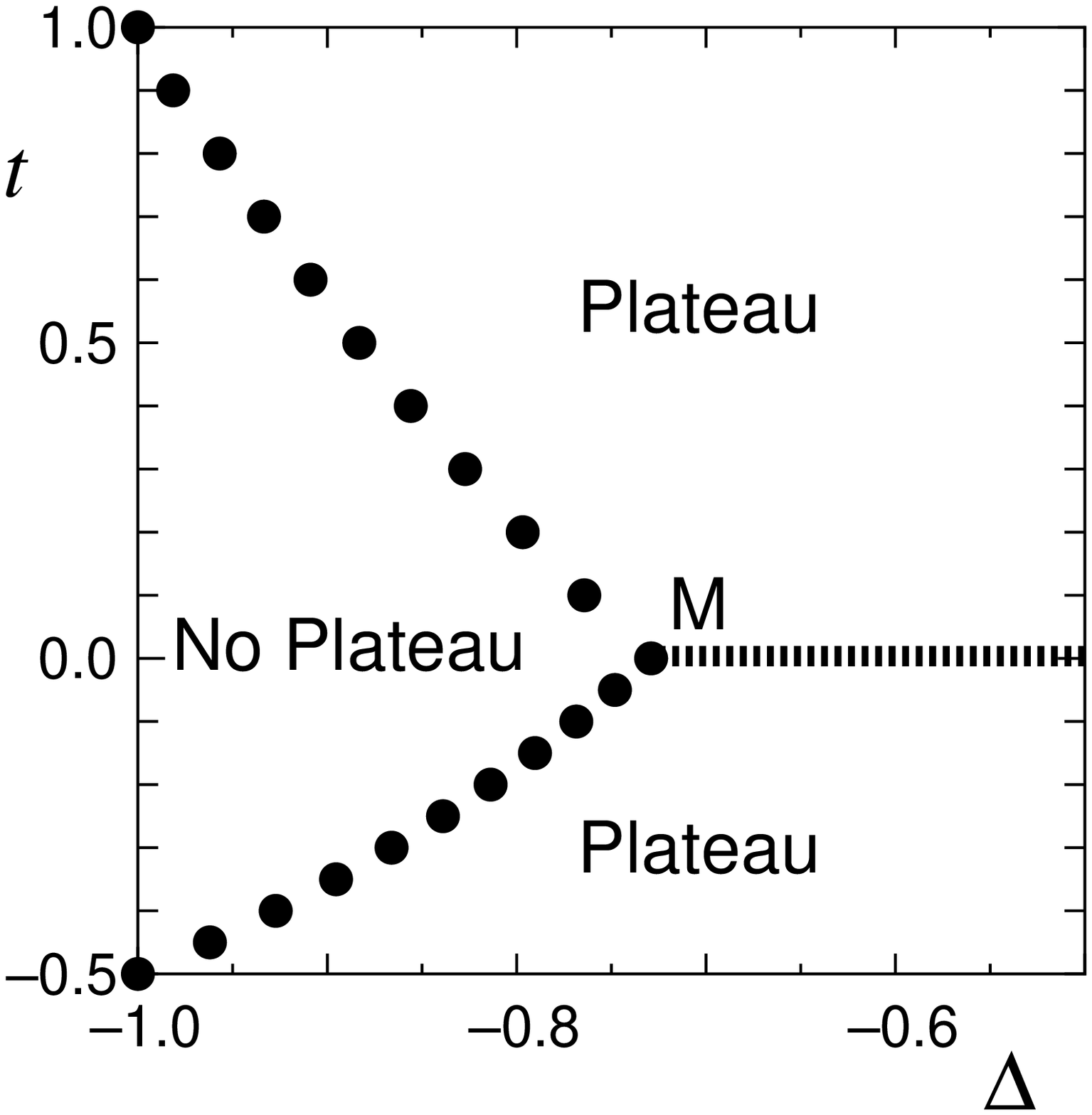}}
      \end{center}
      \caption{Phase diagram on the $\Delta-t$ plane.
               Closed circles are the BKT transition points determined
               from the numerical data as explained in the text.
               Thick dotted line denotes the Gaussian line.
               The multicritical point M is at $(\Delta,t)=(-0.729,0)$.}
      \label{fig:phase-diagram}
   \end{minipage}
\end{figure}

~~~We also investigate the ferromagnetic-ferromagnetic-antiferromagnetic
model [2,4]
\begin{eqnarray}
   H_{\rm FFA}
   &=& \sum_j \left\{
    - J_{\rm F}\left[ h_{3j-2,3j-1}(1)+h_{3j-1,3j}(1) \right] \right. \nonumber \\
   & &~~~~~\left.+J_{\rm A} h_{3j,3j+1}(1)\right\}
\end{eqnarray}
where $J_{\rm F}$ and $J_{\rm A}$ are the ferromagnetic and antiferromagnetic
interaction constants, respectively.
We set $\Delta=1$ so that $H_{\rm FFA}$ is reduced
to the isotropic $S=3/2$ antiferromagnetic chain 
when $\gamma \equiv J_{\rm F}/J_{\rm A} \to \infty$.
Following Hida's numerical calculation, the $M_{\rm s}/3$ plateau clearly
exists when $\gamma \equiv J_{\rm F}/J_{\rm A}$ is small.
On the other hand, it is believed that there exists no 
$M_{\rm s}/3$ plateau in the $S=3/2$ chain.
Thus the plateauful-plateauless transition takes place at finite $\gamma$.
Since the Hamiltonian $H_{\rm FFA}$ is transformed into the generalized version
of the Hamiltonian (1) by the spin rotation [2],
the present method is also applicable to $H_{\rm FFA}$,
resulting in $\Delta_{\rm c}=15.4$ [5].

~~~In conclusion, we have obtained the phase diagram for Hamiltonian (1),
developing a new method to detect the plateauful-plateualess transition
from the numerical diagonalization data.
Our new method is applicable to various systems.

\newpage
\section*{References}

[1] K. Okamoto and A. Kitazawa,
J. Phys. A: Math. Gen. {\bf 32} (1999) 4601.

[2] K. Okamoto,
Solid State Commun. {\bf 98} (1996) 245.

[3] K. Nomura and A. Kitazawa,
J. Phys. A: Math. Gen. {\bf 31} (1998) 7341.

[4] K. Hida,
J. Phys. Soc. Jpn. {\bf 63} (1994) 2359.

[5] A. Kitazawa and K. Okamoto, preprint (1999).

\end{document}